\documentclass[floats,floatfix,showpacs,amssymb,prd,superscriptaddress,twocolumn,aps]{revtex4}
\usepackage{graphicx, epsfig, amssymb} 
\usepackage{amsmath, amsfonts}
\usepackage{bm} 

\usepackage[linktocpage]{hyperref}
\usepackage[caption=false]{subfig}
\usepackage[usenames]{color}

\def\be{\begin{equation}}
\def\ee{\end{equation}}
\def\beq{\begin{eqnarray}}
\def\eeq{\end{eqnarray}}

\newcommand{\bea}{\begin{eqnarray}}
\newcommand{\eea}{\end{eqnarray}}
\newcommand{\ben}{\begin{enumerate}}
\newcommand{\een}{\end{enumerate}}
\newcommand{\bi}{\begin{itemize}}
\newcommand{\ei}{\end{itemize}}

\begin{document}

\title{Wiggly tails: \\ a gravitational
wave signature of massive fields around black holes}

\author{Juan~Carlos~Degollado}
   \affiliation{
   Departamento de F\'\i sica da Universidade de Aveiro and I3N, 
   Campus de Santiago, 3810-183 Aveiro, Portugal.
 }
\affiliation{Departamento de Ciencias Computacionales,
Centro Universitario de Ciencias Exactas e Ingenier\'ia, Universidad de Guadalajara\\
Av. Revoluci\'on 1500, Colonia Ol\'impica C.P. 44430, Guadalajara, Jalisco, M\'exico}

 \author{Carlos~A.~R.~Herdeiro}
   \affiliation{
   Departamento de F\'\i sica da Universidade de Aveiro and I3N, 
   Campus de Santiago, 3810-183 Aveiro, Portugal.
 }


\date{\today}

\begin{abstract}
Massive fields can exist in long-lived configurations around black holes. We examine how the gravitational wave 
signal of a perturbed black hole is affected by such `dirtiness' within linear theory.
 As a concrete example, 
we consider the  gravitational radiation emitted by the infall of a massive scalar field into a
Schwarzschild black hole. Whereas part of the scalar field is absorbed/scattered by the black hole
and triggers gravitational wave emission, another part lingers in long-lived quasi-bound states.
Solving numerically the Teukolsky master equation for gravitational perturbations coupled to the
massive Klein-Gordon equation, we find a characteristic gravitational wave signal, composed by a
quasi-normal ringing followed by a late time tail. In contrast to `clean' black holes, however, the
late time tail contains small amplitude wiggles with the frequency of the dominating quasi-bound
state. Additionally, an observer dependent beating pattern may also be seen. 
These features were already observed in fully non-linear studies; our analysis shows they are
present at linear level, and, since it reduces to a 1+1 dimensional numerical problem, allows for
cleaner numerical data. Moreover, we discuss the power law of the tail and that it only becomes
universal sufficiently far away from the `dirty' black hole. The  \textit{wiggly tails}, by
constrast, are a generic feature that may be used as a smoking gun 
for the presence of massive
fields around black holes, either as a linear cloud or as fully  non-linear hair.
\end{abstract}


\pacs{
11.15.Bt, 
04.30.-w, 
95.30.Sf  
}


\maketitle
\section{Introduction}
\label{sec:intro}
The forthcoming science runs of the second generation gravitational wave (GW) detectors~\cite{Hild:2011np}, and, in parallel, the use of pulsar timing arrays~\cite{Hobbs:2009yy}, are promising a first direct detection of GWs
from astrophysical sources before the decade is over. Such detection will initiate the field of
\textit{GW astrophysics}, a natural realm for  testing general relativity in the strong field
dynamical regime, as well as for constraining theoretical models that predict new gravitational
interactions~\cite{Seoane:2013qna}. This new field should play a significant role over the next decades, delivering ever
increasing precision measurements. Understanding such measurements requires theoretical guidance. As
such, unveiling theoretical GW signatures of physical phenomena, 
and especially of new physics, is particularly timely.

Black holes (BHs) are the most compact objects predicted by general relativity. When involved
in dynamical processes, they play a role as GWs sources. The GW emission pattern from
a `clean' (i.e. vacuum) perturbed BH has long been identified. The BH relaxes back to equilibrium
via damped oscillations in characteristic frequencies -- \textit{quasi-normal ringing}~\cite{Berti:2009kk} --,
determined only by the final equilibrium state. A GW detector far from the BH will not only measure
this ringing but also a late time tail~\cite{Price:1971fb}, due to multiple scatterings by the spacetime curvature of the GWs coming from the
perturbed system. This late time 
tail has therefore the ability to probe the spacetime in the vicinity of the BH.

How does this simple picture change if the BH is `dirty'? That is, if it is surrounded by 
some matter/fields? In a recent thorough examination~\cite{Barausse:2014pra,Barausse:2014tra}, 
it was argued that the quasi-normal ringing will be unchanged, but that resonances could occur in
the late time behaviour of waveforms.  In this paper we shall present a clean resonance which is an effect in GW physics that may be used to identify a specific kind of `dirtiness'.

We consider a simple and tractable model of a dirty environment: a BH 
surrounded by a massive field. The field's mass allows the existence of gravitationally trapped,
long-lived field configurations around the BH, dubbed \textit{quasi-bound states}, characterized by
precise complex frequencies. Actually, massive field configurations can even become infinitely
long-lived in the case of Kerr BHs, for a specific frequency determined by the horizon's angular
velocity~\cite{Hod:2012px,Herdeiro:2014goa,Hod:2013zza,Herdeiro:2014jaa,Herdeiro:2014ima}. The
dynamical interaction of massive fields with BHs
has been mostly discussed in the literature for
scalar fields, both in the test field approximation
(e.g.~\cite{Burko:2004jn,Burt:2011pv,Witek:2012tr}) and in the fully non-linear
regime~\cite{Okawa:2014nda,Guzman:2012jc}. Our study also takes a scalar field as an illustrative
case, but similar statements will
hold for other massive fields, \textit{e.g.}, Proca fields. 

We show that the GW late time tail that follows the quasi-normal ringing of the BH, contains small
amplitude oscillations with the (real part of the) frequency of the dominating
quasi-bound state that endows the BH with a `dirty' environment. 
These oscillations, which were first observed in~\cite{Okawa:2014nda} solving the non-linear Einstein-Klein-Gordon system, are an imprint left by the field's 
cloud on the scatterings that originate the tail. Our setup, based on linear perturbation theory, confirms the generic behaviour described in~\cite{Okawa:2014nda}, and clarifies it is essentially controlled by linear perturbations. Furthermore, the use of 1+1 evolutions allows us to optimize computational resources to extract physical signals. 

The existence of these  \textit{wiggly tails} relies on two ingredients: a perturbed BH that
relaxes by emitting GWs; and a dirty environment around the BH provided by a long lived configuration
of a massive field, with one dominating oscillation frequency. This suggests that these  
tails may be considered as a strong evidence for the existence of massive fields around BHs. 

We also discuss the power law decay of both the GW and scalar field tails. These have known
universal exponents for clean BHs~\cite{lrr-1999-2}. For the scalar case we find a generic decay as
$R^{(s)}\sim t^{-2.5}$ as long as there are no quasi-bound state with significant amplitude at the
extraction point. This is in agreement with the results 
in Ref.~\cite{Burko:2004jn}. Otherwise, the scalar field decay is not a universal power-law any
longer. For the GW tail, we find that, as long as the extraction point is sufficiently far away from
the `dirty' BH, it decays as $R^{(g)}\sim t^{-1.5}$. This is a different behaviour from `clean' BHs,
for which the Price decay law is $t^{-(2\ell+3)}$ with $\ell=2$~\cite{Price:1971fb}. Otherwise, the
GW decay is affected
by the dirtiness and no generic behaviour could be unveiled. 

This paper is organized as follows: after this introduction  
we describe the scalar field in the background of a BH in section \ref{sec:model}. In section
\ref{sec:perturbations} we introduce the linear
gravitational perturbation equations in terms of the Weyl scalar $\Psi_4$ as well as the
harmonic decomposition to get a set of 1+1 coupled partial differential equations. In section
\ref{sec:numerics} we describe the numerical code used to solve them and the numerical results.
Finally, in section
\ref{sec:conclusions} we give some concluding remaks.


\section{A model for a `dirty'  perturbed black hole}
\label{sec:model}
%
As a simple model of a perturbed BH in a `dirty' environment we shall consider the interaction of a 
Schwarzschild BH with a scalar field $\Phi$, with mass $\mu$
and  stress-energy tensor
\begin{equation}
 T_{\mu\nu} = \Phi_{,\mu}\Phi_{,\nu}-\frac{1}{2}g_{\mu\nu}\left(\Phi^{,\sigma}\Phi_{,\sigma}+
\mu^2\Phi^2 \right)\ .
\label{eq:tmunu}
\end{equation}
The conservation of the stress-energy tensor implies that the field obeys
the Klein-Gordon equation $\Box \Phi=\mu^2\Phi$. We write the background geometry in ingoing Kerr-Schild
coordinates, which are horizon penetrating and hence more adequate for a numerical treatment. The corresponding line element is
\begin{equation}
 ds^{2} =- f(r)dt^2 + \frac{4M}{r}drdt + \left( 1+\frac{2M}{r} 
\right)dr^2 \\  +r^{2} d\Omega_2 \ ,
\label{eq:ds_ks}
\end{equation}
where  $f(r)\equiv \left(1-\frac{2M}{r}\right) $ and $d\Omega_2$ is the standard line element on $S^2$.

In order to find the solutions of interest of the Klein Gordon equation, the scalar field $\Phi$ can be expanded
in Fourier modes of frequency $\omega$ and spherical harmonics $Y_0^{\ell,m}$ with spin weight zero.
Requiring that the field is ingoing at the horizon and approaches zero at spatial infinity leads to
a discrete set of complex frequencies, for each $\ell$, corresponding to a fundamental mode and
overtones. These solutions are the quasi-bound states. These states have well known 
frequencies of oscillation and rates of decay, \textit{cf.} for instance
\cite{Detweiler:1980uk,Zouros:1979iw,Dolan:2007mj}, which depend on the value of
$\mu$. The trend to keep in mind, for the
Schwarzschild background, is that increasing the mass of the scalar field implies decreasing the life-time
of the quasi-bound states, which is measured by the (inverse of the) imaginary part of the complex frequencies. For very small $\mu$, \textit{e.g} $10^{-23}$ eV, a value compatible with some scalar field dark matter
models \cite{Matos:2000ss}, quasi-bound states may last for cosmological
time scales around a  supermassive BH, say with $\sim 10^8 M_\odot$, without being significantly absorbed~\cite{Burt:2011pv}. 
In such cases, the field configuration can be considered as a continuous source of perturbation. 
Below we shall give specific frequencies for the cases of interest herein, obtained using the continued
fraction method first described in~\cite{Leaver:1985ax}. 

In this paper, we study the  gravitational perturbations of the background
sourced by the scalar field. These perturbations are evolved simultaneously with the Klein-Gordon equation in the
unperturbed background, since the background perturbations lead to second order effects in the
Klein-Gordon equation.
The model is the following. We perturb the geometry by setting, at some initial time, a wave
packet of the scalar field in the background (\ref{eq:ds_ks}). 
A generic wave packet contains a range of frequencies that includes quasi-bound state frequencies but
also contains enough dynamics to awake the quasi-normal modes of the BH.
Thus, in the evolution of the scalar field wave packet, there is a part which lingers
around the BH in one dominating long-lived quasi-bound state. Other sub-leading quasi-bound states
are also present, and lead to one observable effect, as we shall describe. Here, long-lived
means that the time scale for the decay of the quasi-bound state is much longer than the time
scale for the quasi-normal ringing.

\section{Scalar field sourced gravitational perturbations}
\label{sec:perturbations}
%
To compute the GWs induced by the backreaction of the scalar field on the metric we use linear 
perturbation theory in the Newman-Penrose formalism. The radiative information  is extracted
from the Newman-Penrose scalar 
 $\Psi_4 = -C_{\alpha\beta\gamma\delta}k^{\alpha}m^{*\beta} k^{\gamma} m^{*\delta}$,
where $C_{\alpha\beta\gamma\delta}$ is the first order perturbed Weyl tensor, the vectors $(l^{\alpha} , m^{\alpha} , k^{\alpha} )$ are elements of a Newman-Penrose
null basis \cite{Newman62a,Chandrasekhar83} and $m^{*\alpha}$ is the complex conjugate of
$m^{\alpha}$. 

For distant observers, $\Psi_4$ describes outgoing GWs. A central observation, due to 
Teukolsky~\cite{Teukolsky:1972my,Teukolsky:1973ha}, is that for (background) spacetimes of Petrov
type D the first order perturbation of $\Psi_4$ decouples from the perturbations of the other
Newman-Penrose scalars. Furthermore, he showed that the resulting wave equation  can be separated in
the Kerr (and thus Schwarzschild) background in its angular and radial parts, by expanding the
perturbation in a basis of spheroidal harmonics with the appropriate spin weight.

To solve numerically the radial perturbation equation coupled to the Klein-Gordon equation
we use the line element (\ref{eq:ds_ks}) and the null vector basis $k^{\mu} = (1,-1,0,0)$,  
\begin{eqnarray*}
 l^{\mu} = \frac{1}{2}\left(2-f(r),f(r),0,0\right) , \ \ \  m^{\mu} = \frac{1}{\sqrt{2r}}( 0,0,1,i\csc\theta) \ .
\end{eqnarray*}
The details of the derivation of the equation for the perturbed Weyl scalar $\Psi_4$ are given in
\cite{Nunez:2011ej}. After decomposing $\Psi_4$  in terms of spin weighted spherical harmonics (with weight -2):
\begin{equation}
\Psi_4=\frac{1}{r}\sum\limits_{\ell,m}\,R^{(g)}_{\ell,m}(t,r)\,{Y_{-2}}^{\ell,m}(\theta,\varphi) \ ,
\end{equation}
the radial-temporal perturbation equation reads
 \begin{eqnarray} \label{eq:psi_pi_g}
&-\left(1+\frac{2M}{r}\right)\,\partial_{t t}R^{(g)}_{\ell,m}
 +  \left(1-\frac{2M}{r}\right)\,\partial_{ r r}R^{(g)}_{\ell,m} 
+\frac{4M}{r}\,\partial_{ r t }R^{(g)}_{\ell,m}
&\nonumber
\\
& + 2\,\left(\frac{2}{r}+
\frac{M}{r^2}\right)\,\partial_{ t}R^{(g)}_{\ell,m} 
+ 2\,\left(\frac{2}{r}
-\frac{M}{r^2}\right)\,\partial_{ r}R^{(g)}_{\ell,m}
& \nonumber
\\
&
+
\left(\frac{2M}{r^3} -\frac{(\ell-1)(\ell+2)}{r^2} \right)R^{(g)}_{\ell,m} = 16\pi r S(R^{(s)}_{\ell,m}) \ ,
\end{eqnarray}
where the sources $S(R^{(s)}_{\ell,m})$, are projections of the stress-energy tensor
(\ref{eq:tmunu}) along
the null tetrad; see for example Eqs. (3.10)-(3.21) of Ref. \cite{Nunez:2011ej}.  



To solve the second order equation \eqref{eq:psi_pi_g} we decompose it into a
first order system using the first order variables
\begin{equation}
\psi_{\ell,m}^{(g)}\equiv \partial_r R^{(g)}_{\ell,m}\ , \ \   \ \
\Pi_{\ell,m}^{(g)}\equiv \frac{r+2\,M}r\,\partial_t\,
R^{(g)}_{\ell,m} -2\,\frac{M}r\,\psi_{\ell,m}^{(g)} \ .
 \label{eq:Pig}
\end{equation}
Then, the following first order equations are obtained:
\begin{eqnarray}
&&\partial_t\,R^{(g)}_{\ell,m}=\frac{1}{r+2\,M}\left(r\,\Pi_{\ell,m}^{(g)} + 2\,M\,\psi_{\ell,m}^{(g)}
\right) \ ,
\label{eq:evolR}\\
&&\partial_t\,\psi_{\ell,m}^{(g)}=\partial_r\left[\frac{1}{r+2\,M}\left(r\,\Pi_{\ell,m}^{(g)} +
2\,M\,\psi_{\ell,m}^{(g)}
\right)\right], \, \label{eq:evolPsi} \\
&&\partial_t\,\Pi_{\ell,m}^{(g)}=\frac{1}{r+2\,M}\left(2\,M\,\partial_r\,\Pi_{\ell,m}^{(g)}
+r\,\partial_r\,\psi_{\ell,m}^{(g)} \right)  \nonumber \\
&& + \frac{2}{r\,(r+2\,M)^2}\left[\left(2\,r^2+5\,M\,r+4\,M^2\right)\,\Pi_{\ell,m}^{(g)}  \right.
\nonumber
\\
&&+ \left. \left(r+4\,M\right)\,\left(2\,r+3\,M\right)\psi_{\ell,m}^{(g)} \right] \nonumber \\
&&+ \left(2\frac{M}{r^3} -
\frac{\left(\ell-1\right)\,\left(\ell+2\right)}{r^2}\right)\,R^{(g)}_{\ell,m} -
16\pi r S(R^{(s)}_{\ell,m}) \ .
\label{eq:evolPig}
\end{eqnarray}

To compute the evolution of the source in (\ref{eq:evolPig}), we must follow the dynamics of 
the scalar field,  given by the Klein-Gordon equation. 
We write
this equation in the coordinates \eqref{eq:ds_ks} and decompose the field as
\begin{equation}
 \Phi=R^{(s)}_{\ell,m}(t,r)\,{Y_{0}}^{\ell,m}(\theta,\varphi) \ .
\end{equation}
Then a second order partial differential equation for $R^{(s)}_{\ell,m}(t,r)$ is obtained.
In order to solve it, we perform a first order decomposition, in close
analogy with that performed for the gravitational perturbations, by defining $\psi_{\ell,m}^{(s)}\equiv \partial_r
R^{(s)}_{\ell,m}$
and 
\begin{equation}
\Pi_{\ell,m}^{(s)}\equiv \frac{r+2\,M}r\,\partial_t\,
R^{(s)}_{\ell,m} -2\,\frac{M}r\,\psi_{\ell,m}^{(s)} \ .
 \label{eq:Piesca}
\end{equation}
With these first order variables the Klein-Gordon equation becomes an evolution equation for
$\Pi^{(s)}_{\ell,m}$,
\begin{eqnarray}
 \partial_{t}\Pi^{(s)}_{\ell,m} &=& \frac{1}{r+2M}\left( 2M\partial_r \Pi^{(s)}_{\ell,m}+r\partial_r
\psi^{(s)}_{\ell,m}  \right) \nonumber \\ \nonumber
&&+\frac{2}{r(r+2M)^2}(\psi^{(s)}_{\ell,m}-\Pi^{(s)}_{\ell,m} )\\ 
&&+ \left(
\mu^2-\frac{\ell(\ell+1)}{r^2}+\frac{2M}{r^3} \right)R^{(s)}_{\ell,m} \ .
\label{scalar2}
\end{eqnarray}

With the solution of the scalar field at each time step we reconstruct the stress-energy tensor of 
the scalar field \eqref{eq:tmunu} and then the source term in (\ref{eq:evolPig}). 


\section{Numerical results}
\label{sec:numerics}
%
We numerically solved the coupled system (\ref{eq:evolR})--(\ref{eq:evolPig}) 
and (\ref{eq:Piesca})--(\ref{scalar2}), for the evolution of the gravitational perturbations and of
the scalar field.  We have used the
method of lines with a third order Runge Kutta time integration and finite
differencing with sixth order stencils. Such accuracy is needed in order to capture correctly the
late time tail behaviour. The use of horizon
penetrating coordinates allow us to set the inner
boundary inside the horizon.
The outer boundary is typically located at $r=1000M$ (although in some cases we set it
at $r=2000M$ in order to avoid any contamination coming in from the boundary). At the last point of 
the numerical grid we set up the incoming modes to zero.
%


We chose as initial data a static Gaussian perturbation of the scalar field, with the form
$R_{1,0}^{(s)}(0,r)=e^{-(r-r_g)^2/\sigma^2}$. For a typical run, the pulse was centred at 
$r_g=10M$ and $\sigma=0.5M$. The gravitational perturbation is initially set to zero, since we are
interested in the
gravitational waveform produced
as a response to the presence of the scalar field. This choice of initial data does not bias the
result.  
It has been observed, by numerically evolving arbitrary initial data, that the appearance of
quasi-bound states is generic \cite{Barranco:2012qs}. This guarantees that our
choice of initial data will create the `dirty' environment we seek and simultaneously excite the quasi-normal ringing of the BH; but
other choices would also achieve the same goal. 
%
%
%
%
Indeed, we observe that part of the scalar field falls
into the BH, part is
scattered and another part remains as quasi-bound states.

\begin{figure}[h!]
\begin{center}
\includegraphics[width=0.48\textwidth]{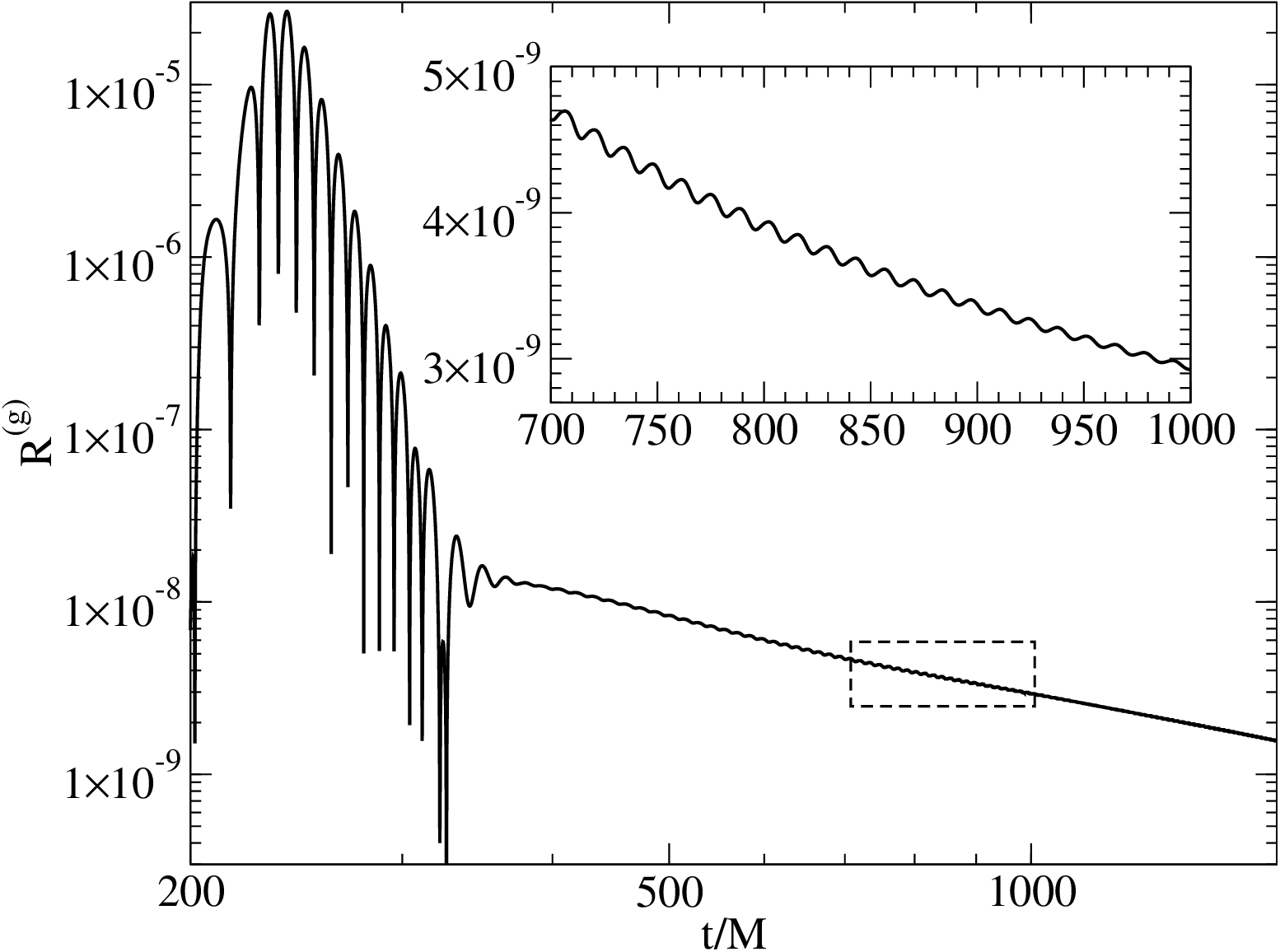}
\caption{GW signal ($\ell=2$) for $M\mu =0.48$  (observer at $500M$). Small amplitude wiggles can be seen in
the late time behaviour, in an otherwise apparently power law tail.}
\label{wiggyt}
\end{center}
\end{figure}

Fig.~\ref{wiggyt} exhibits the gravitational signal for $M\mu =0.48$, as measured by a
sufficiently distant observer. 
As expected, the quasi-normal ringing is 
followed by a late time tail. We have found that the dirtiness 
is negligible for the ringing, in agreement with the results
in~\cite{Barausse:2014pra,Barausse:2014tra}: 
performing a sinusoidal fit of the first stages of the signal, we obtained that the quasi-normal ringing has the frequency of the quadrupolar ($\ell=2$) gravitational
mode $\omega^{QNM}=0.373-\textit{i}\ 0.0889$ of a clean Schwarzschild BH
\cite{lrr-1999-2,Berti:2009kk,Konoplya:2011qq}.

A careful inspection of the late time behaviour, on the
other hand, 
unveils a novel feature: 
on top of the 
decaying behaviour there are oscillations with a very small amplitude -- Fig.~\ref{wiggyt} (inset).
The frequency of the oscillations in this \textit{wiggly tail} precisely correspond to the real part
of the frequency of the dominating, \textit{i.e.} largest amplitude, quasi-bound state. In this
case the dominating state is the  first overtone, with frequency $\omega_1 = 0.4717 - \textit{i}\
1.4501\times10^{-3}$ \cite{Burt:2011pv}.

\begin{figure}[h!]
\begin{center}
\includegraphics[width=0.5\textwidth]{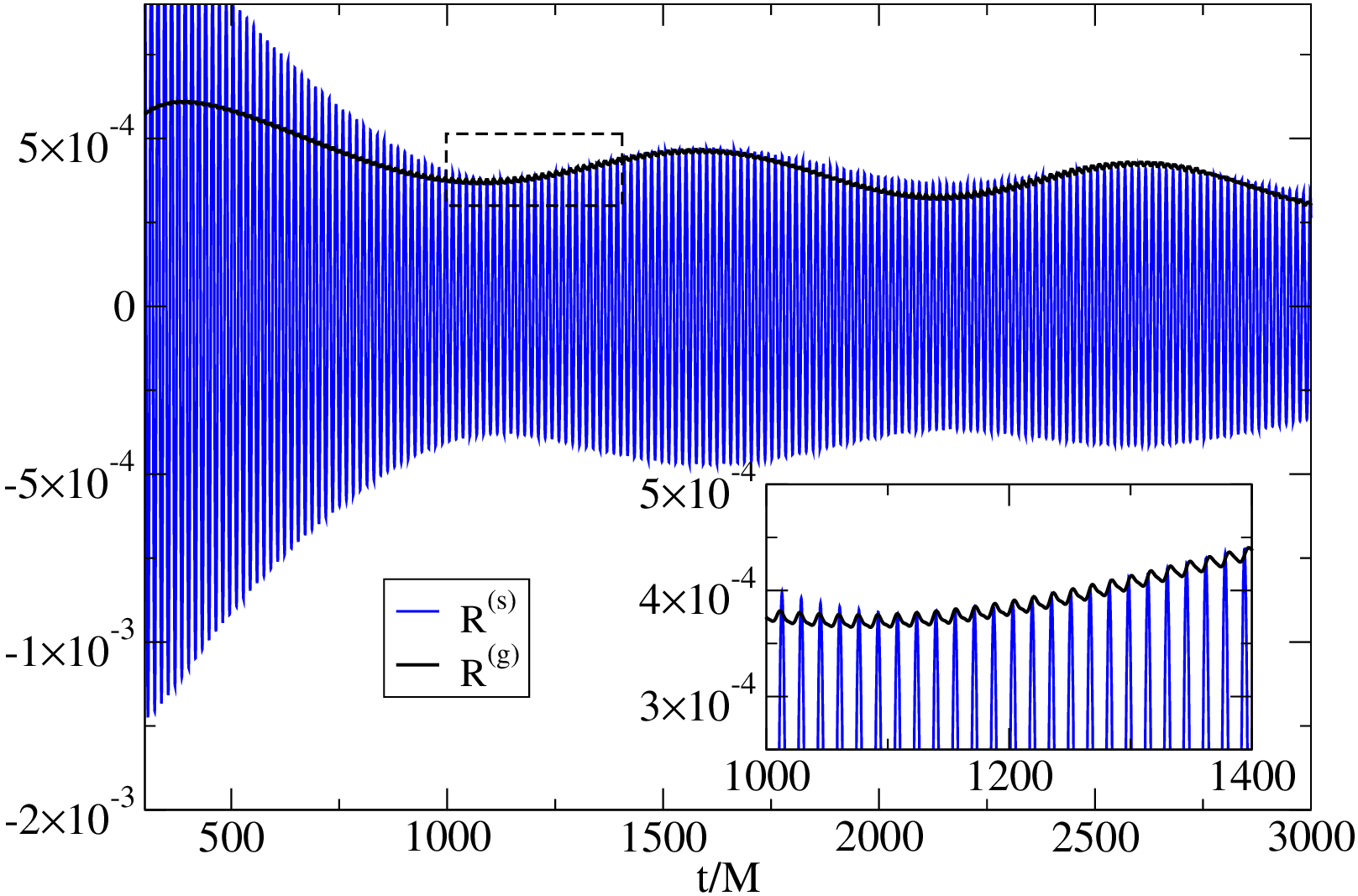}
\caption{Scalar ($\ell=1$) and GW ($\ell=2$) signal for $M\mu = 0.40$ (observer at $70M$). 
The dominant frequency of
the scalar field is that of the first quasi-bound state overtone, with 
$\omega_1 = 0.3955 - \textit{i}\  2.2668 \times10^{-4}  $; the next leading quasi-bound state is the second overtone, with 
$\omega_2 = 0.3975 - \textit{i}\  1.0035\times10^{-4} $. }
\label{fig:grav_scalbeat}
\end{center}
\end{figure}

To clearly demonstrate the observation in the last paragraph and simultaneously exhibit yet another effect that may be present in the late time
signal, we present a slightly different value of the field's mass $M\mu =0.4$, for which we monitor
the behaviour of both the scalar field and of the GW signal -- Fig.
\ref{fig:grav_scalbeat}. 
In this figure, the GW signal has been rescaled for better visualization. 
The first observation is that the late time GW signal (black solid line) presents, besides the 
aforementioned wiggles, a \textit{beating} pattern. The figure shows that the beating in the GW
signal is resonating the beating observed in the scalar field. Beating patterns are typical in
systems with two dominating frequencies with comparable amplitudes. In this case, the beating
frequency corresponds to the difference between the first and
second overtones.
The second observation is that the frequency of the wiggles seen in the late time GW signal
coincides with the frequency of the dominating quasi-bound state(s). This is shown in the figure's 
inset. 

\begin{figure}[h!]
\begin{center}
\includegraphics[width=0.5\textwidth]{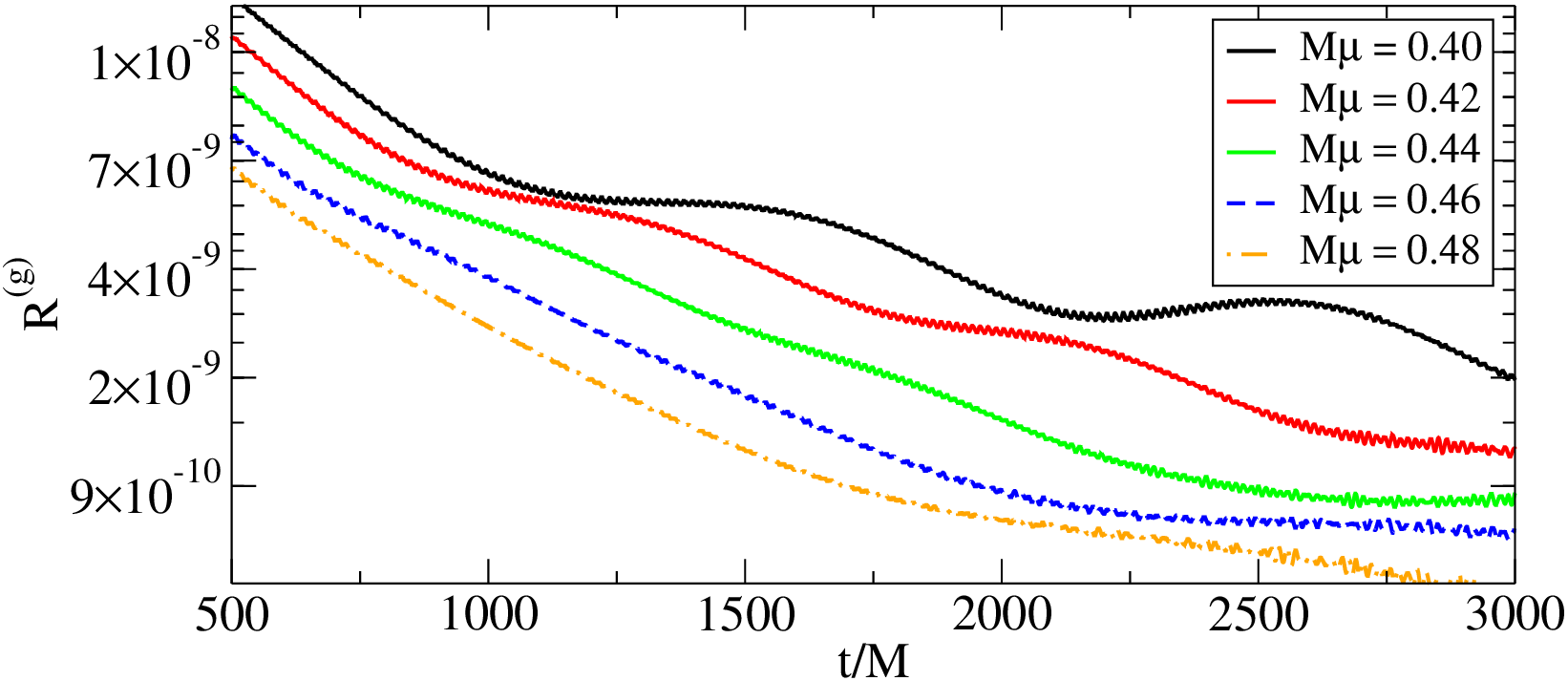}
\caption{Late time behaviour of the GW signal ($\ell=2$) for several
values of $M\mu$ (observer at $70M$). The beating becomes increasingly more visible as the mass is
decreased from $M\mu=0.48$ to $M\mu=0.4$. For even smaller values the beating becomes again
suppressed
(not shown).}
\label{fig:beatingGW}
\end{center}
\end{figure}

Both the reported effects seen in the late time tails -- the wiggles and the beating -- depend on
the value of the field's mass $\mu$. For the wiggles, the mass yielding the largest amplitude is
$M\mu \simeq 0.48$. One may think of this optimal mass as a balance between two effects. Smaller
$\mu$ increases the life-time of the `dirtiness'; however, it also increases its spatial dispersion,
and thus decreases the local effect of the perturbation. For the beating, the dominating quasi-bound
states should have comparable amplitudes
and the difference in frequencies should yield a period compatible with the observation times. An
empirical conclusion, for our sets of initial data, is that the beating is most visible for $M\mu
\simeq 0.4$ -- Fig.
\ref{fig:beatingGW}.

The existence of beating patterns for scalar fields around BHs has been discussed previously in
different models~\cite{Witek:2012tr,Degollado:2013bha,Okawa:2014nda}. But the study herein first
reports a GW signal clearly resonating with such beating. 
We notice, however, that the beating -- unlike the wiggles -- depends on the extraction radius. For
instance, if the observation point is close to any of the nodes of one of the states contributing to the
beating, this effect vanishes both in the scalar field and in the GW counterpart. Furthermore, the
beating is sensitive to the initial data, since, although quasi-bound states are excited for generic
initial data, the amplitudes of these states depend on the details of such data. 

\begin{figure}[h!]
\begin{center}
\includegraphics[width=0.48\textwidth]{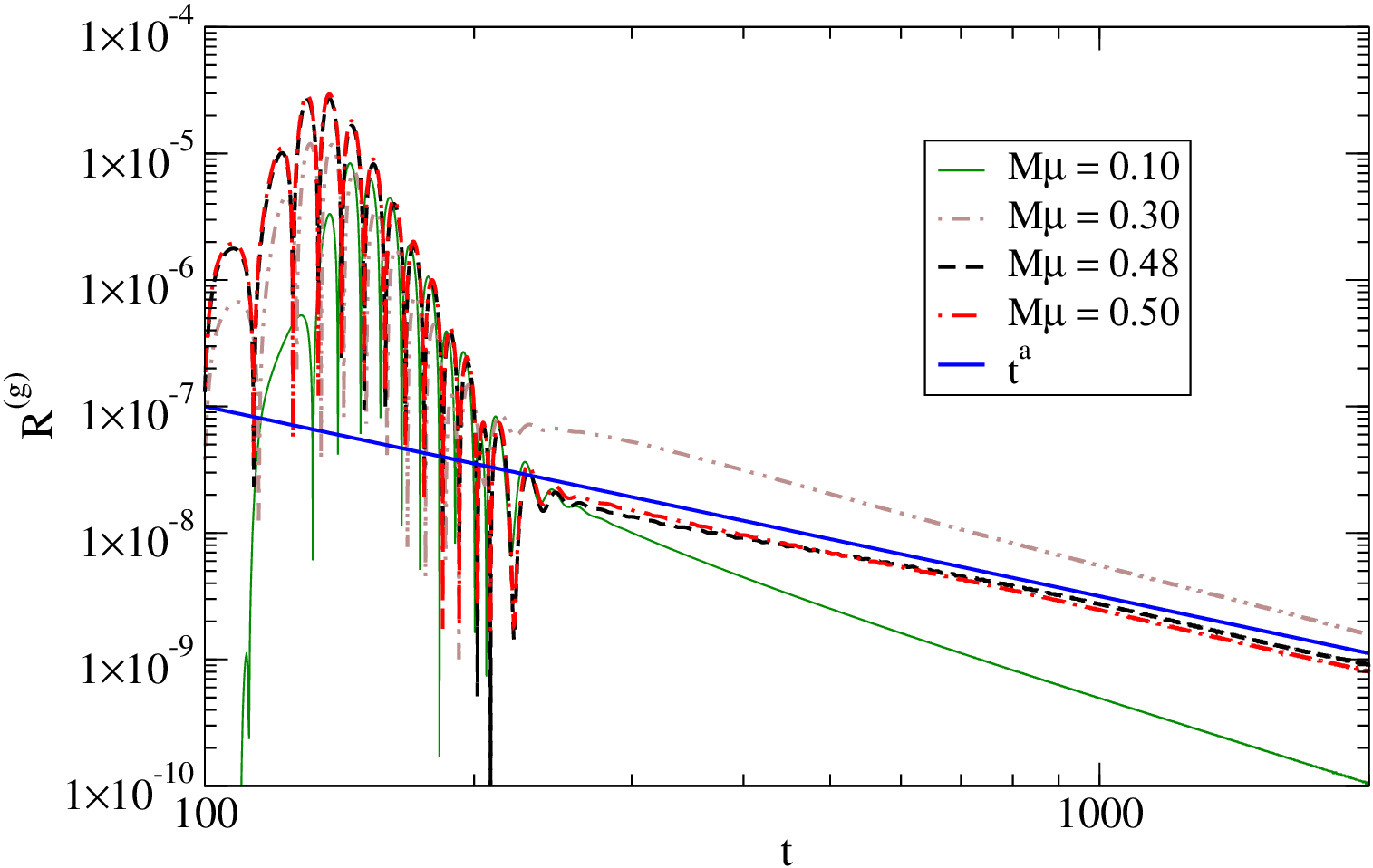}
\caption{Quadrupolar GW signal (observer at $100M$). We show a power law fit $t^a$ with $a=1.5$ 
for the late time tail. This power law fits well the three curves with highest mass, but it fails for the lowest mass.}
\label{wiggyt2}
\end{center}
\end{figure}

Finally, we have observed that the late time behaviour of the GW signal has a power 
law when extracted sufficiently far away from the `dirty' BH - Fig. \ref{wiggyt2}. If the extraction takes place within the `dirtiness', this is not the case, as illustrated by the curve with $M\mu=0.1$ in Fig.  \ref{wiggyt2}, since for this value of the mass, the dominant quasi-bound state has a significant amplitude at the extraction position.

\section{Concluding remarks}
\label{sec:conclusions}
%
%
%
%
We have shown that the presence of a massive field in long-lived configurations around a BH leaves a distinctive signature in the late time behaviour of the GW signal when the BH is perturbed: a wiggly tail. 
The frequency of these wiggles is determined by the mass of the BH. For stellar mass BHs, $M\sim 1 M_\odot - 10 M_\odot$, this frequency lies in the range $\omega\sim 12.7$ kHz - 1.27 Hz, which, as for the corresponding quasi-normal modes, falls in the bandwidth of ground base detectors such as aLIGO~\cite{Harry:2010zz}. For supermassive BHs, on the other hand,  $M\sim 10^6 M_\odot - 10^9 M_\odot$, $\omega\sim 12.7$ mHz - $12.7 \mu $Hz, entering the bandwidth of the space based detectors such as eLISA~\cite{Seoane:2013qna}. The amplitude of the wiggles, however, is $10^4$ times smaller than that of the corresponding quasi-normal modes. As such the detection of such a signal will be a considerable technical challenge.

Concerning the mass of the scalar field, the values considered here $M\mu\sim 0.4-0.5$, correspond to $\mu\sim  10^{-17}-10^{-20}$ eV, for supermassive BHs and  $\mu\sim  10^{-11}-10^{-12}$ eV for stellar mass BHs. Whereas these values are extremely small when compared to the masses of standard model particles, such light particles have been suggested in the context of high energy physics scenarios beyond the standard model. For instance, they arise naturally in string compactifications, \textit{i.e.} the axiverse~\cite{Arvanitaki:2009fg}.


Both the existence of wiggles and of a beating in the GW tails are qualitative features that may be used to distinguish a dirty environment from an isolated BH. Interestingly, the mere presence of a beating is significant, since it provides evidence for two comparable amplitude frequencies in the dirty environment. For instance, for a Kerr BH with scalar hair~\cite{Herdeiro:2014goa}, there should be
a clearly dominating frequency -- that of the background scalar field -- and hence no noticeable
beating should occur. 

The observation of any of these features provides a quantitative measure of the frequencies of the quasi-bound states involved in the `dirtiness' and hence of the mass of the field surrounding the BH.

\section*{Acknowledgements}
We would like to thank Vitor Cardoso for useful comments on a draft of this paper.
This work was partially supported by the NRHEPÐ295189
FP7-PEOPLE-2011-IRSES Grant. J.C.D. acknowledges support from FCT via project No.
PTDC/FIS/116625/2010. 


\newpage

\bibliography{num-rel}

 
\end{document}